\documentclass[aip,graphicx,reprint]{revtex4-1}
\usepackage{graphicx}
\usepackage{float}
\usepackage{epsfig,psfrag}
\usepackage{amsmath, amsthm, amssymb}
\draft 

\begin{document}

\title[25~kHz narrow spectral bandwidth of a wavelength tunable diode laser with a short waveguide-based external cavity]{25~kHz narrow spectral bandwidth of a wavelength tunable diode laser with a short waveguide-based external cavity}

\author{R.M. Oldenbeuving}
\email{R.M.Oldenbeuving@UTwente.nl}
\affiliation{University of Twente, Laser Physics and Nonlinear Optics group, P.O.~Box~217, 7500~AE, Enschede, The Netherlands}
\affiliation{MESA+ Research Institute for Nanotechnology, P.O.~Box~217, 7500~AE, Enschede, The Netherlands}

\author{E.J. Klein}
\affiliation{XiO Photonics, P.O.~Box~1254, 7500~BG, Enschede, The Netherlands}

\author{H.L. Offerhaus}
\affiliation{MESA+ Research Institute for Nanotechnology, P.O.~Box~217, 7500~AE, Enschede, The Netherlands}
\affiliation{University of Twente, Optical Sciences group, P.O.~Box~217, 7500~AE, Enschede, The Netherlands}

\author{C.J. Lee}
\affiliation{University of Twente, Laser Physics and Nonlinear Optics group, P.O.~Box~217, 7500~AE, Enschede, The Netherlands}	
\affiliation{MESA+ Research Institute for Nanotechnology, P.O.~Box~217, 7500~AE, Enschede, The Netherlands}
\affiliation{FOM Institute DIFFER, Edisonbaan~14, 3439~MN, Nieuwegein, The Netherlands}

\author{H. Song}
\affiliation{Delft University of Technology, Delft Center for Systems and Control, Mekelweg~2, 2628~CD, Delft, The Netherlands}
\affiliation{Zhejiang~University, Institute of Underwater Technology and Ship Engineering, Yuhangtang~Road~866, 310058, Hangzhou, China}

\author{K.-J. Boller}
\affiliation{University of Twente, Laser Physics and Nonlinear Optics group, P.O.~Box~217, 7500~AE, Enschede, The Netherlands}
\affiliation{MESA+ Research Institute for Nanotechnology, P.O.~Box~217, 7500~AE, Enschede, The Netherlands}

\date{\today}

\begin{abstract}
We report on the spectral properties of a diode laser with a tunable external cavity in integrated optics. Even though the external cavity is short compared to other small-bandwidth external cavity lasers, the spectral bandwidth of this tunable laser is as small as 25~kHz (FWHM), at a side-mode suppression ratio (SMSR) of 50~dB. Our laser is also able to access preset wavelengths in as little as 200~$\mu$s and able to tune over the full telecom C-band (1530~nm -- 1565~nm).
\end{abstract}


\maketitle

\section{Introduction}
\label{Introduction}

Tunable narrow bandwidth diode lasers with bandwidths well below 1~MHz are of interest for applications in, e.g., coherent optical communications, where frequency tunability and narrow bandwidths can be used to increase data transfer density~\cite{Ip2008}. These lasers might also be of interest in applications where increased phase stability is required, such as atomic clocks~\cite{Jiang2011}. Another application of interest is broadband, squint-free K$_u$-band phased array antenna systems, where the narrow laser bandwidth increases the resolution and where laser wavelength tuning is necessary for phase retrieval~\cite{Marpaung2011,Meijerink2010}. Here the beat frequency between a fixed frequency and a tunable laser, with sufficiently narrow bandwidths, provides phase information needed to replace complicated electrical delay lines and modulators in current designs. Free-space external cavity diode lasers are capable of producing such narrow bandwidths, however wavelength tuning is too slow, due to a moving grating or feedback mirror~\cite{Saliba2009}. 

The spectral bandwidth of free-running diode lasers without frequency selective feedback is in the order of GHz, so spectral narrowing methods have to be applied in order to reach bandwidths well below 1~MHz. Such narrowing, in combination with spectral tuning can be reached through tunable frequency selective optical feedback~\cite{Henry1982}. The two main approaches are monolithic semiconductor lasers, such as distributed Bragg reflector (DBR) lasers and distributed feedback (DFB) lasers, and external cavity semiconductor lasers (ECSL). However, the bandwidth of DBR and DFB lasers are typically around a few~MHz with a tuning range of several~nm, though tunability can be increased at the expense of bandwidth~\cite{Signoret2004,Laue2001}. DBR and DFB lasers are not suitable if a narrower bandwidth combined with wider tunability is required. To solve this problem, external cavities can be used~\cite{Saliba2009,Henry1982,Signoret2004,Laue2001}. However, to reach bandwidths well below 1~MHz, fluctuations in the optical cavity length have to be avoided, so mechanically stable free space external laser cavities with lenses, gratings, and lengths of at least several centimeters are required. It is difficult to achieve mechanically stable cavities without any acoustic noise. As an example, the Schawlow-Townes limit~\cite{Henry1982} of these lasers can be reached by active control of the positioning of the focusing lenses inside the cavity very precisely using piezo elements~\cite{Saliba2009}.In addition, the problem in achieving both a narrow bandwidth and a fast wavelength tunability, is the mechanical stability of the external cavity of the diode laser; on the one hand, fast tuning capabilities will result in a less mechanically stable cavity and hence a bandwidth broadening, on the other hand, a very mechanically stable cavity will result in lower tuning speed and tuning range.

We present a solution to these problems that follows the approach of Chu~\emph{et~al.}~\cite{Chu2009,Chu2010}, where an optical gain chip is coupled to an external cavity that is integrated on a waveguide chip. We call this approach a waveguide based external cavity semiconductor laser (WECSL). The external cavity provides frequency selective, and tunable feedback to the laser diode. The waveguide chip incorporates a double micro-ring resonator (MRR) structure, as depicted in Fig.~\ref{FigWaveguideLayout}, referred to as an MRR mirror. The resonance frequencies of both MRRs can be tuned by heating the MRRs, resulting in faster tunability than possible for mechanically tuning a free-space external cavity. Since the external cavity is integrated on a waveguide chip, mechanical stability is significantly increased compared to free-space external cavities.

While Chu~\emph{et~al.} --through this approach-- had presented a side-mode suppression ratio (SMSR) of 40~dB and a wide wavelength coverage (38~nm around 1550~nm), we have found an improved SMSR of 50~dB and demonstrate rapid wavelength addressability. Moreover, Chu~\emph{et~al.} make no mention of their achieved spectral bandwidth, while, here, we demonstrate a spectral bandwidth of 25~kHz, which is surprisingly narrow considering the relatively small physical length of the WECSL.

In this paper we will discuss the design, fabrication, performance and spectral properties of our WECSL. We show the measured SMSR, tuning range, switching time, and demonstrate a very narrow spectral bandwidth. The experimentally obtained spectral bandwidth is compared to the theoretically expected value.

\section{Waveguide design}
\label{WaveguideDesign}
The spectral gain bandwidth of the diode laser gain chip used in our experiments extends over a significant range of several tens of nanometers. This may easily lead to undesired oscillation at several frequencies and also decrease the SMSR. A careful choise of the MRRs free spectral range (FSR) allowed oscillation to be restricted to a single mode with a narrow bandwidth, with high SMSR, while also achieving wide wavelength tunability. More specifically, the FSR of the MRRs were chosen such that feedback over most of the gain bandwidth, except for a selected wavelength of laser operation, was avoided. In addition to the FSR, the strength of the in- and output coupling of the MRR, through its coupling coefficient ($\kappa$), must also be chosen.

The free spectral range, $\Delta\lambda_{FSR}$, of a single MRR, as given by~\cite{Yariv2000}, can be calculated by 

\begin{equation}
\Delta\lambda_{FSR} = \frac{\lambda^2}{n_{g}L}
    \label{Eq:FSR}
\end{equation}

where $\lambda$ is the central wavelength, $n_{g}$ is the group index of the waveguide and $L$ is the geometrical roundtrip length of the MRR. By changing the optical roundtrip length $n_{g}\cdot L$, the resonant frequencies are shifted. In our case, the group index is varied by controlled heating of the MRR via an electric heater placed atop the MRR. Using this method of tuning, for the MRRs used in this work, with L in the order of 350~$\mu$m, one is capable of shifting the resonant frequencies over the full range of the FSR. 

To extend the tuning range beyond one FSR, a second MRR with a slightly different radius can be added in series, exploiting the Vernier effect~\cite{Geuzebroek2002,Grover2002}, rendering an FSR of the combined MRRs, $\Delta\lambda_{FSR,tot}$, of 

\begin{equation}
 \left.\begin{aligned}
\Delta\lambda_{FSR,tot} & = m \cdot \Delta\lambda_{FSR,1}\\
\Delta\lambda_{FSR,tot} & = (m+1) \cdot \Delta\lambda_{FSR,2}
       \end{aligned}
 \right\}
 \label{Eq:FSRvernier}
\end{equation}

where $\Delta\lambda_{FSR,1}$ and $\Delta\lambda_{FSR,2}$ are the free spectral ranges of the rings with radii $R_1$ and $R_2$, respectively, and $m$ is the smallest positive integer that simultaneously satisfies Eq~(\ref{Eq:FSRvernier}). Note that Eq.~(\ref{Eq:FSRvernier}) does not take the spectral width of the transmission resonances of the MRRs into account, but only at their center wavelengths. The finite width of the transmission resonance can be taken into account by using Eq.~(\ref{Eq:FWHM},\ref{Eq:Lorentzian}), with their respective constraints.

The spectral bandwidth at full width half maximum (FWHM), $\Delta\lambda_{FWHM}$ of the resonance of a single MRR, as given by Yariv~\cite{Yariv2000}, is described by

\begin{equation}
\Delta\lambda_{FWHM}=\frac{\lambda_0^2}{\pi L n_{g}}\cdot \frac{\kappa^2}{\sqrt{1-\kappa^2}}
    \label{Eq:FWHM}
\end{equation}

where $\lambda_0$ is the center wavelength of the MRR resonance, and $\kappa$ is the coupling coefficient of the MRR (for the electric-field amplitude). Eq.~(\ref{Eq:FWHM}) only incorporates the MRR's roundtrip losses via $\kappa$, while neglecting other losses (such as scattering, absorption and radiative bending losses). This is a valid approximation for the MRRs used in this research, since --for the typical loss values discussed below-- the coupling coefficient of the MRR accounts for over 99\% of the roundtrip losses.

In designing the double MRR feedback, such that the laser oscillates only at a single frequency, it is important that $R_{\lambda_{0}}/R_{\lambda_{sp}} > G_{\lambda_{0}}/G_{\lambda_{sp}}$, where $R_{\lambda_{0,sp}}$ and $G_{\lambda_{0,sp}}$ are the reflectivities and relative gains at the ratio of the feedback at the wavelength that provides the highest feedback (called ``center peak'') with that of the next peak in the reflectivity spectrum (called ``side peak''), respectively. The ``peak-height ratio'' (PHR) is given by $R_{\lambda_{sp}}/R_{\lambda_{0}}$. $R_{\lambda_{sp}}$ is dependent on the coupling coefficients, $\kappa$, of the MRRs (i.e., on the FWHM of the MRRs) and on the difference between the FSRs of the two rings. More specifically, a calculation of the PHR requires taking the spectral shape of the resonances of the individual MRRs into account. For simplicity, we have only considered the case for a single side peak. However, for completeness, it should be noted that -- if the FWHM of the two MRRs are assumed equal -- the first two side peaks occur at \mbox{$\lambda_{sp}=\lambda_0 \pm (|\Delta\lambda_{FSR,1}-\Delta\lambda_{FSR,2}|/2)$}. It is well known (see e.g. Little~\emph{et~al.}~\cite{Little1997}) that the shape of the transmission resonances of a single MRR can be approximated by a Lorentzian shape

\begin{equation}
L(\lambda)=\left(\frac{1}{\pi}\right)\frac{\frac{1}{2} \Delta\lambda_{FWHM}}{(\lambda-\lambda_c)^2+(\frac{1}{2} \Delta\lambda_{FWHM})^2}
    \label{Eq:Lorentzian}
\end{equation}
 
were $L(\lambda)$ is the Lorentzian function and \mbox{$\lambda_c=\lambda_0 \pm n\cdot \Delta\lambda_{FSR}$} (with n an integer). To calculate the feedback of two MRRs coupled in series, as we investigate here, the center peak occurs at \mbox{$\lambda_c=\lambda_0$} and the side peak occurs at $\lambda_{sp}$. The power transmitted through both rings is now described by the product of the Lorentzian transmission functions of the two individual MRRs at \mbox{$\lambda_c=\lambda_0$} and the power of the side peaks can be approximated by the product of the Lorentzian functions \mbox{$L(\lambda_{sp})$ with $\lambda_c=\lambda_0 \pm \Delta\lambda_{FSR,1}$}. Assuming equal coupling coefficients $\kappa$ for the MRRs, the PHR ($\delta_{PHR}$) is obtained as 

\begin{equation}
\delta_{PHR}=\frac{L(\lambda_{sp})^2}{L(\lambda_c)^2}.
\label{Eq:PHR}
\end{equation}

Summarizing, with Eqs.~(\ref{Eq:FSR}-\ref{Eq:PHR}) a double-MRR mirror can be characterized with a $\Delta\lambda_{FSR,tot}$, a FWHM and a PHR. A double MRR that provides a sufficiently large $\Delta\lambda_{FSR,tot}$, a small FWHM, and a low PHR, should result in a narrow bandwidth laser, operating on a single-longitudinal mode, with a high SMSR, and a wide tuning range.

\begin{figure}[t]
\vbox{\hbox to\hsize{\psfig{figure=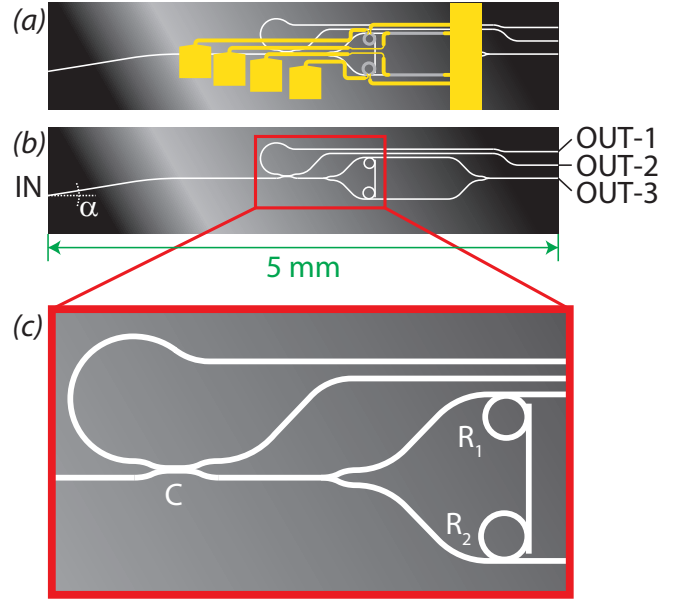,width=\hsize,clip=}\hfill}}
\caption{Schematic of the waveguide chip. (a) Shows the complete waveguide chip. The waveguides are depicted in white, electrical contacts are in yellow, and heaters are in gray. The heaters are placed on top of the MRRs (for thermal tuning of the MRRs' resonance frequencies), and on the straight waveguides after the MRR mirror (for changing the refractive index of the straight waveguides after the MRRs, such that a maximum output power can be achieved after the two straights are combined). (b) Shows the waveguide chip without heaters and electrical contacts.  (c) Is a zoom-in on the two MRRs, having radii $R_1$ and $R_2$, and the coupler $C$.}
\label{FigWaveguideLayout}
\end{figure}

We have fabricated such frequency selective waveguide circuits, based on a double MRR, using box-shaped Si$_3$N$_4$/SiO$_2$ waveguide technology (TriPleX$^{TM}$)~\cite{Morichetti2007}, because this offers low waveguide scattering losses ($\leq$0.06~dB/cm), a relatively high index contrast ($\Delta$n=0.1-0.5) and low radiative bending losses ($\sim$ 1~dB/cm for a bend with a radius of 50~$\mu$m)~\cite{Morichetti2007,Heideman2009}. The box-shaped design of the waveguide cross section used in this research has a mode-field diameter (MFD) of $\sim$1.5~$\mu$m. A radius of R=50~$\mu$m for MRRs is the smallest radius that can be fabricated reproducibly, with acceptable radiative bending losses (we chose 1~dB/cm to be acceptable, because bending losses $<$1~dB/cm can be safely neglected with respect to in- and output coupling as described above). With a designed $n_{g}$=1.73, and $R_1$=50~$\mu$m, gives $\Delta\lambda_{FSR,1}$=4.4~nm at $\lambda_0$=1550~nm. For applications in telecommunications, it is required to be able to tune over at least the full C-band (1530~nm -- 1565~nm), i.e., a $\Delta\lambda_{FSR,tot}>$35~nm. To allow for an error margin in the FSR, we choose to aim for an $\Delta\lambda_{FSR,tot}>$40~nm.  Eq.~(\ref{Eq:FSRvernier}) then requires $m\geqslant$10 which leads to a $\Delta\lambda_{FSR,tot}$=44~nm and $\Delta\lambda_{FSR,2}$=4.0~nm (such that the design diameter of the second MRR is $R_2$=55~$\mu$m). The layout of the frequency selective mirror, consisting of Y-splitters, directional and bi-directional couplers, and MRRs, is shown in Fig.~\ref{FigWaveguideLayout}. The bi-directional symmetric coupler taps off $\sim$5\% of the light from position $C$ into what we call a measurement channel. This feature was included to allow the laser and mirror performance to be monitored in more detail. The coupler is located between the laser diode input port and the MRR mirror, see Fig.~(\ref{FigWaveguideLayout}b,c). \emph{OUT-1} and \emph{OUT-2} indicate the output ports of the measurement channel and \emph{OUT-3} indicates the output port of the WECSL. Finally, to optimize coupling between the gain section, which is an anti-reflection coated angled single-stripe diode gain chip, and the external waveguide cavity, the input of the waveguide is oriented at an angle ($\alpha \approx 9^\circ$) with respect to the waveguide chip facet. This angle takes into account the output angle of the gain chip ($\approx 5^\circ$) and the different refractive indices of the gain chip and waveguide.

Since the PHR needs to be smaller than the steepest slope in the gain profile of the laser, a sufficiently small PHR needs to be provided by design. The steepest slope of the gain spectrum of the laser gain chip used in this research is 0.48~dB/nm, requiring that $\delta_{PHR}<$0.63 (i.e., $>$2.0~dB over $\lambda_0-\lambda_{sp}$). Combining Eqs.~(\ref{Eq:Lorentzian} and~\ref{Eq:PHR}), with the parameters determined earlier ($R_1$=50~$\mu$m, $R_2$=55~$\mu$m, and $n_{g}$=1.73) shows that $\Delta\lambda_{FWHM}$=0.52~nm, requiring $\kappa^2<$0.44.

\begin{figure}[t]
\vbox{\hbox to\hsize{\psfig{figure=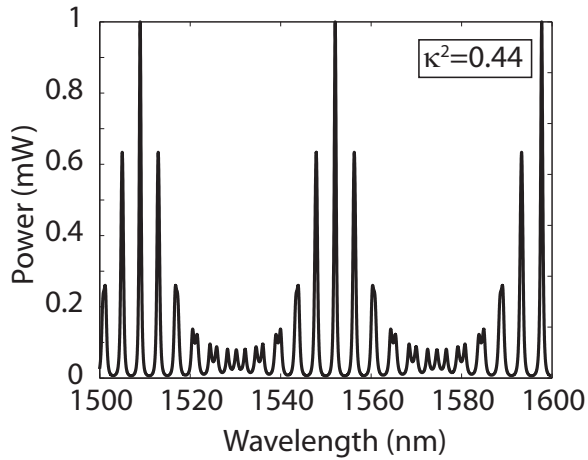,width=\hsize,clip=}\hfill}}
\caption{Calculated reflectance spectrum of the MRR mirror, using $\kappa^2$=0.44, R$_1$=50~$\mu$m, R$_2$=55~$\mu$m and n$_{g}$=1.73, with no additional roundtrip losses.}
\label{FigVernierMRRMirrorResponse}
\end{figure}

To validate these results, we have calculated the values for the FSR, FWHM, and PHR in more detail, using the general method, as described by Yariv~\cite{Yariv2000,Yariv2002}. A very detailed description of this method can be found in the PhD Thesis of Klein~\cite{Klein2007}. These calculations resulted in a set of optimized parameters in agreement with those found above ($\Delta\lambda_{FSR,tot}$=44~nm, $\delta_{PHR}$=0.63, and $\Delta\lambda_{FWHM}$=0.51~nm, for $\kappa^2$=0.44). The calculated reflectivity spectrum, taking R$_1$=50~$\mu$m, R$_2$=55~$\mu$m and n$_{g}$=1.73, is shown in Fig.~\ref{FigVernierMRRMirrorResponse}.

Due to the many different steps in fabrication of the used waveguides, it is very challenging to fabricate \mbox{--purely by design--} an MRR mirror with the exact coupling coefficient of $\kappa^2$=0.44. In practice, the intended coupling coefficient is expected to possess a fabrication error of at least \mbox{$\kappa^2 \pm 0.05$}. To make, nevertheless, a suitable value of $\kappa$ available, we fabricated a large set of MRR mirrors in which the nominal design value of $\kappa$ is systematically varied in steps smaller than the fabrication error. In our case we have designed the MRR mirrors such that the values for $\kappa$ cover at least 0.3$<\kappa^2<$0.4. Thereby we ensure that, in spite of fabrication errors, an MRR mirror with a $\delta_{PHR}<$0.63 can be obtained. In addition, this makes MRR mirrors available with a large variety of $\kappa$ values, which are of interest for investigating variations in the performance of MRR mirrors.

\section{Waveguide characterization}
\label{WaveguideCharacterization}

For the measurements described in Sections \ref{WaveguideDesign}, \ref{WaveguideCharacterization}, \ref{LaserCharacterization}, and \ref{WECSLcharacterization}, we used an optical spectrum analyzer (OSA) (ANDO AQ6317), a super luminescent diode (SLD) (Thorlabs S5FC1005S), a calibrated fiber-coupled power meter (HP81536A), a calibrated beam profiler (Thorlabs BP104-IR), a custom anti-reflection coated C-band angled-stripe gain chip (Fraunhofer Heinrich-Hertz-Institut), an analog output card (National Instruments PCIe-6251), a 20~GHz optical detector (Discovery Semi DSC30s), an acousto-optic modulator (AOM) (AA Opto Electronic, MT80-IIR60-F10-PM0.5-130.s), an AOM driver (ISOMET 232A-2), an Erbium doped fiber amplifier (Firmstein Technologies Inc. PR25R) and an RF spectrum analyzer (Agilent Signal Analyzer MXA N9020A). 

\begin{figure}[t]
\vbox{\hbox to\hsize{\psfig{figure=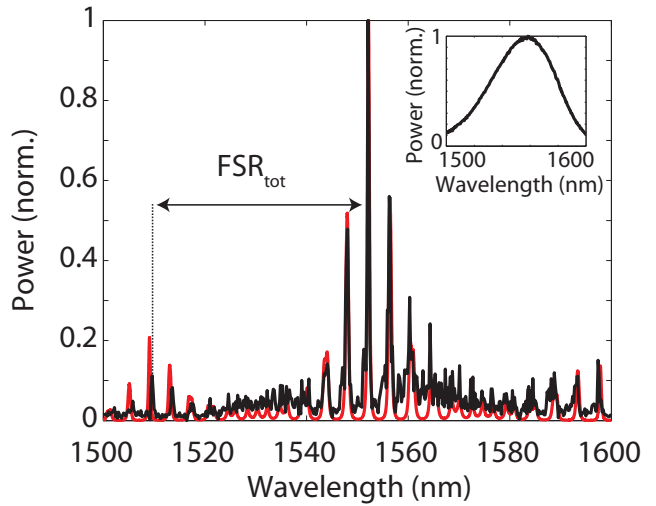,width=\hsize,clip=}\hfill}}
\caption{The fitted calculated (red line) and measured (black line) reflected spectra of the MRR mirror, normalized to the maximum power of the reflected spectra. The output spectrum of the SLD (inset) was used as the input spectrum for the MRR mirror for both the measurement and calculation. For the calculations $\kappa^2$=0.39, and n$_{g}$=1.7293 was used. The PHR was measured to be 0.56 and the $\Delta\lambda_{FSR,tot}$ was measured to be 42.6~nm.}
\label{FigReflectedSpectrum}
\end{figure}

We characterized a set of ten MRR mirrors with different designed $\kappa$-values. As expected, we found several MRR mirrors with a measured value of 0.3$<\kappa^2<$0.4. For characterization in more detail, out of this set of MRR mirrors, we selected the MRR mirror that showed the highest output when coupled to a single-mode fiber. All waveguide characterization and laser characterization measurements presented in this letter are from that MRR mirror. The spectrally dependent reflection from the MRR mirror was measured using the fiber coupled super luminescent diode, butt coupled via a PM-fiber to the input port of the waveguide chip (Fig.~\ref{FigWaveguideLayout}(b), \emph{IN}). The spectrum of the SLD is shown in the inset in Fig.~\ref{FigReflectedSpectrum}, normalized to the maximum spectral power density of the SLD. The spectrum of the light reflected by the MRR mirror is measured at the output port of the measurement channel (Fig.~\ref{FigWaveguideLayout}(b), \emph{OUT-1}) using the OSA. This is shown in Fig.~\ref{FigReflectedSpectrum}, normalized to the maximum spectral power density of the reflected spectrum.

The peak power measured at port~\emph{OUT-1} is about 20\% of the peak power measured at port~\emph{OUT-2}, which indicates that the MRR mirror has an overall reflectivity of 20\% at the center wavelength $\lambda_0$=1552.2~nm. For other measured MRR mirrors this value was typically between 20\% and 30\%. 
The measured peak-height ratio is 0.56$\pm$0.01. When compensated for the normalized input spectrum of the SLD this corresponds to a $\delta_{PHR}$ of 0.55$\pm$0.01. The theoretical spectrum calculated as described above, was fit to the measured spectrum in Fig.~\ref{FigReflectedSpectrum}. To fit the wavelength of the calculated center peak to the measured center peak at $\lambda_0$ at 1552.2~nm the group index n$_{g}$ was varied in the calculations. For fitting the PHR, the coupling coefficient $\kappa$ was varied. The fit yielded a value of n$_{g}$=1.7293$\pm$0.0001, which is in good agreement with the value expected by the design parameters n$_{g}\approx$1.73. The value of $\delta_{PHR}$=0.55 obtained by the fit corresponds to $\kappa^2$=0.39$\pm$0.01, which is in the designed range of 0.3$<\kappa^2<$0.4. As can be seen in Fig.~\ref{FigReflectedSpectrum}, the measured $\Delta\lambda_{FSR,tot}$ of the MRR mirror is 42.6~nm, which is very close to the designed $\Delta\lambda_{FSR,tot}$ of 44.0~nm. The residual deviations can be addressed to deviations in ring radii from the design values.

\section{Free-running gain chip characterization}
\label{LaserCharacterization}

The free-running gain chip was characterized to compare the spectral properties --laser threshold, maximum output power, side-mode suppression ratio (SMSR) and spectral bandwidth--  of the laser gain chip with and without external cavity.

The gain chip has a specified typical back-facet reflectivity of 85\% and maximum front facet reflectivity of 0.1\%. The latter is achieved by an anti-reflection coating and an angle of 5$^{\circ}$ of the gain stripe to the surface normal of the output facet. The specified far-field divergence is 19$^{\circ}$ FWHM in both directions, which corresponds to a mode-field diameter at the output facet of 6~$\mu$m FWHM. The specified threshold pump current is 17.0~mA. 

We verified the threshold pump current to be I$_{th}$=17.8~mA. The maximum output power was measured to be 7.4~mW at I$_{LD}$=90.0~mA. At a pump current well above threshold (I$_{LD}$=30~mA) the observed side-mode suppression ratio (SMSR) was -24~dB. At higher pump currents (I$_{LD}>$60~mA) the SMSR gradually reduced to -20~dB.
 
Spectral measurements were performed using a single-mode fiber connection to the input of the OSA to ensure maximum spectral resolution. The spectral bandwidth was measured to be 0.02~nm FWHM (i.e., 2.5~GHz). Because this value is higher than the resolution of the OSA (0.01~nm, i.e., 1.25~GHz), we can assume that the spectral bandwidth of the free-running laser was resolved by the OSA. 

\section{WECSL characterization}
\label{WECSLcharacterization}

The gain chip, described in Section~\ref{LaserCharacterization}, was butt-coupled to the MRR mirror, described in Sections~\ref{WaveguideDesign}~and~\ref{WaveguideCharacterization}, to form a WECSL.

Due to the feedback of the MRR mirror, the threshold pump current was lowered to a value of I$_{th}$=5.6~mA. 

To accurately measure the output power, the out-coupled light was collected using a multimode fiber. The maximum output power was measured, at a pump current of I$_{LD}$=86.7~mA, to be 1.0~mW~$\pm$4.5$\cdot$10$^{-2}$~mW. Increased feedback by the MRR mirror, in place of the weak feedback from the residual diode facet reflectivity, is expected to increase the output, especially as the threshold was also measured to be lowered. Our observation of a lower output power can be explained by loss due to a mode mismatch between the gain chip and the waveguide. Additionally, we expect losses by reflections at the facets of the waveguide chip and the multi-mode fiber. 

\begin{figure}[t]
\vbox{\hbox to\hsize{\psfig{figure=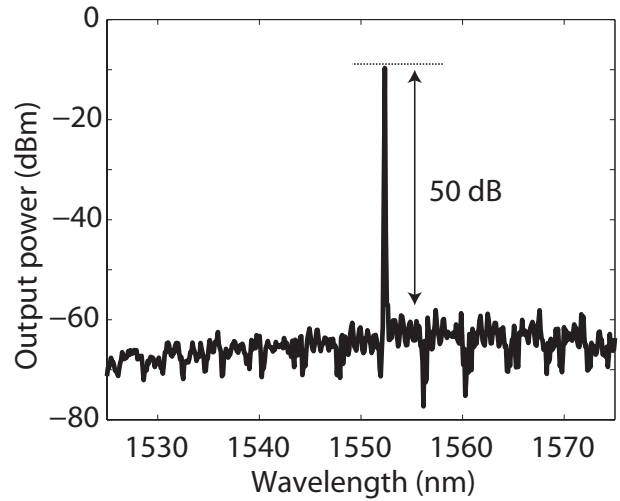,width=\hsize,clip=}\hfill}}
\caption{Output power spectrum of the WECSL on a logarithmic scale. The highest measured side-mode suppression ratio is 50~dB.}
\label{FigSMSR}
\end{figure}

The output spectrum, measured across the range of 1525~nm to 1575~nm, is displayed in Fig.~\ref{FigSMSR}. It can be seen that, compared to the free-running case, the maximum value for the SMSR (at a pump current of I$_{LD}$=86.1~mA) drastically improved to 50~dB. Depending on the alignment of the gain chip to the MRR mirror's $IN$-port, the SMSR was found to vary between typically 40~dB and 50~dB. It should be noted that in Fig.~\ref{FigSMSR} the maximum power is lower than the previously mentioned 1.0~mW (i.e., 0~dBm). This can be addressed to the mode mismatch of the MFD of the waveguide output (\emph{OUT-3} in Fig.~\ref{FigWaveguideLayout}b) with the MFD of the single-mode fiber.

\begin{figure}[t]
\vbox{\hbox to\hsize{\psfig{figure=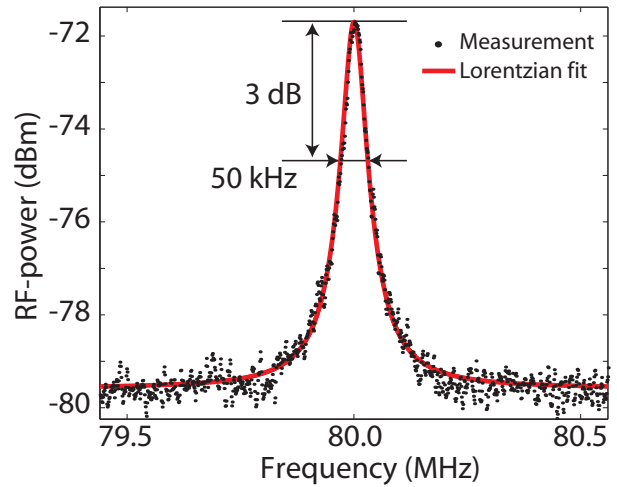,width=\hsize,clip=}\hfill}}
\caption{Self-heterodyne beat spectrum the WECSL. The black dots show the measured RF beat spectrum and the red line shows a Lorentzian fit with a 3~dB bandwidth of 50~kHz. This corresponds to a laser bandwidth of 25~kHz.}
\label{FigSpectralLineWidth}
\end{figure}

The OSA had inadequate spectral resolution to resolve the spectral bandwidth of the WECSL. To measure the spectral bandwidth of the laser with significantly increased resolution, we used a delayed self-heterodyne interferometer as described in~\cite{Tsuchida1990,Han2005} with a delay line of 16~km of single-mode fiber. The measured power spectrum of the RF-beat signal is shown in Fig.~\ref{FigSpectralLineWidth}, where the black dots indicate the measurement data and the red line shows a Lorentzian fit. The Lorentzian fit has a 3~dB bandwidth of 50~kHz, which corresponds to a 25~kHz FWHM laser bandwidth. The phase noise of the RF oscillator is in the Hz regime, so that does not significantly contribute to the measured bandwidth of the laser. 

This laser bandwidth of 25~kHz is surprisingly narrow, when only the physical length of the external waveguide cavity (several mm) and the maximum reflectance of the MRR mirror are taken into account. However, the narrow bandwidth can, indeed, be explained by the following two physical effects, which are introduced by the specific design of the double-MRR mirror. First the MRRs --because the resonant light performs multiple roundtrips-- introduce an optical delay~\cite{Poon2004}. This results in an increase of the cavity photon lifetime, due to the relatively high quality factors (Q factors) of the MRRs. The Q factor of a resonator, such as a Fabry Perot or an MRR, is given by $Q=\lambda/(\Delta\lambda_{FWHM})$~\cite{Driessen2005}. The cavity photon lifetime (sometimes referred to as cavity ring-down time) is given by $\tau_{c}=\lambda Q/(2\pi c)$~\cite{Driessen2005}. For a Fabry Perot (FP) cavity, it is convenient to rewrite the cavity photon lifetime to directly include the effective cavity length ($L$), the reflectance of the front mirror ($R_a$) and end mirror ($R_b$) and the cavity-internal loss ($Loss$). $\tau_c$ then becomes $\tau_{c,FP}=-2L/(c\cdot ln(R_a R_b (1-Loss)^2))$. Assuming the WECSL can be described as a FP laser, $R_a$ can be taken as the reflection of the back mirror of the gain chip and $R_b$ as the maximum reflectivity of the MRR mirror, the photon cavity lifetime of the WECSL is the sum of the individual cavity lifetimes. With two MRRs inside the cavity, the overall $\tau_c$ becomes $\tau_{c,FP}+\tau_{c,MRR_1}+\tau_{c,MRR_2}$ (where $\tau_{c,MRR_{1,2}}$ is the cavity photon lifetime of the first and second MRR, respectively). Using the values as described in Section~\ref{WaveguideDesign}~and~\ref{WaveguideCharacterization}, $\tau_c$ is calculated to be 13.5~ps.

Second, unlike the diode chip back facet that provides a spectrally broadband reflection, the MRR mirror provides a frequency selective feedback. This narrows the laser bandwidth even further, which can be viewed as due to chirp reduction~\cite{Kazarinov1987} and optical self-locking~\cite{Laurent1989}. Using Eq.(\ref{Eq:Bandwidth}) as given by~\cite{Laurent1989}, one can calculate the minimum expected bandwidth of the laser, $\Delta\nu$, as

\begin{equation}
\Delta\nu=\frac{\Delta\nu_{S.T.}}{\left(\frac{n_1 L_1}{n_0 L_0}\cdot\frac{F_1}{F_0}\right)^2},
\label{Eq:Bandwidth}
\end{equation}

where $n$ the index of refraction, $L$ the physical length, and $F$ the Finesse, with subscripts 0: the gain chip part of the WECSL and subscripts 1: the waveguide part of the WECSL. $\Delta\nu_{S.T.}$ is the Schawlow Townes bandwidth of the free running laser diode ($\Delta\nu_{S.T.}=\Delta\nu_0/(1+\alpha)^2$)~\cite{Laurent1989}, where $\Delta\nu_0$ is the FP cavity bandwidth of the free running laser diode and $\alpha$ the line width enhancement factor of the free running laser diode~\cite{Henry1982}, in our case, approximated as $\alpha=5$, which is a good approximation since, typically, 4$<\alpha<$8~\cite{Saliba2009,Henry1982}. The Finesse is calculated as $F=\Delta\lambda_{FSR}/(\Delta\lambda_{FWHM})$~\cite{Driessen2005}. For the free running laser diode, we calculated $F_0$=0.89 and for the external cavity, using the cavity photon lifetime to calculate the FWHM, and $\Delta\lambda_{FSR,tot}$, as measured in Section~\ref{WaveguideCharacterization}, we calculated $F_1$=440. 

Taking the optical delay and frequency selective feedback into account, we calculate the theoretical minimum bandwidth of our WECSL to be as narrow as 5~Hz. The calculations show that the minimum bandwidth could be reduced even further by designing a new MRR mirror with three major improvements. First, the cavity photon lifetime can be increased through decreasing $\kappa$ of the MRR. Second, by decreasing the cavity-internal loss by mode matching the waveguide to the gain chip, the cavity photon lifetime increases. Finally, by increasing the $\Delta\lambda_{FSR,tot}$, the Finesse increases.

In the experiments reported here, this theoretical limit is not achieved. We believe that there are two main explanations. First, a mode mismatch in coupling between the gain chip and the waveguide leads to increased intra cavity losses. Second, we believe that acoustic vibrations of the residual gap between gain chip and waveguide chip cause the cavity loss and lenght to fluctuate, both of which lead to spectral broadening (via amplitude modulation in the case of cavity loss and increased phase noise in the case of length modulation).

Nevertheless, a theoretical minimum laser bandwidth of 5~Hz clearly shows the huge potential of such WECSLs, as future light sources with extreme spectral purity. In particular, because this potential is much higher than expected from standard, grating-controlled, external cavity diode lasers. 

\begin{figure}[t]
\vbox{\hbox to\hsize{\psfig{figure=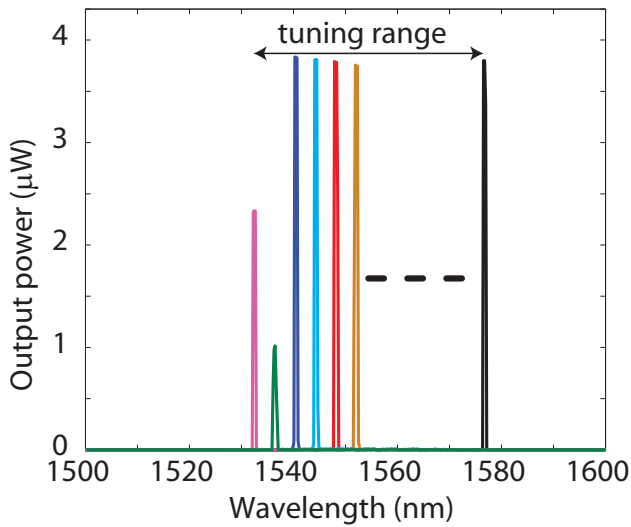,width=\hsize,clip=}\hfill}}
\caption{Superposition of several output spectra of the WECSL as obtained when increasing the heater voltage of a single MRR (with radius $R_1$). The overall spectral coverage, indicated as tuning range, is over 44~nm.}
\label{FigTuningRange}
\end{figure}

The full tuning range of our laser was measured to be at least 44~nm, as can be seen in Fig.~\ref{FigTuningRange}. The tuning range covers the entire telecommunication C-band (1535~nm -- 1560~nm). 

\begin{figure}[t]
\vbox{\hbox to\hsize{\psfig{figure=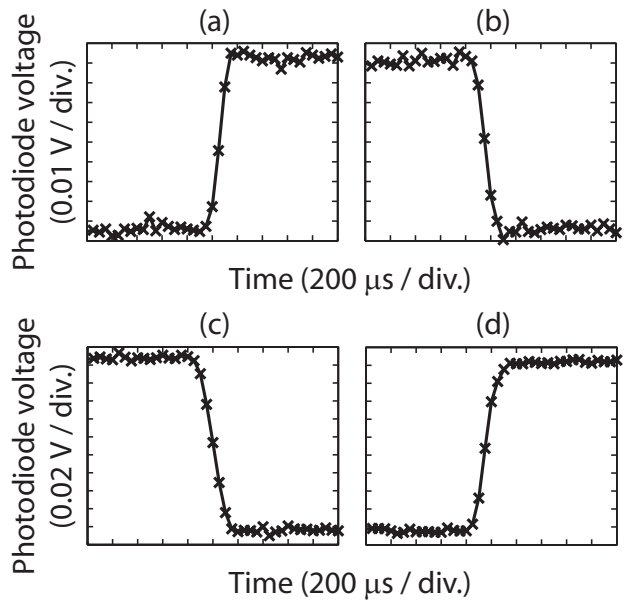,width=\hsize,clip=}\hfill}}
\caption{Measured laser switching times. In (a) and (b) the photodiode is aligned to detect light with a wavelength of 1552.2~nm, and in (c) and (d) the photodiode is aligned to detect light with a wavelength of 1548.2~nm. In (a) and (c) the laser is switched from 1548.2~nm to 1552.2~nm, whereas in (b) and (d) the laser is switched from 1552.2~nm to 1548.2~nm.}
\label{FigTuningSpeed}
\end{figure}

To quantify the minimum time in which the laser wavelength can be set to a different value within the tuning range, we measured what we define as the switching time between two preset wavelengths. To induce an adjustable switching time between two preset wavelengths, we used a computer controlled voltage supply, connected to a home-built power amplifier to drive the heaters. The output ports from the power amplifier were connected to the electrical contacts on the waveguide chip. By applying a voltage to only the heater of the ring with $R_1$=50$\mu$m, we expect to see wavelength switches of the size of the FSR of $R_2$ (i.e., 4.0~nm). Indeed, the first stable and reproducible wavelength switch we observed occurred from $\lambda$=1552.2~nm to $\lambda$=1548.2~nm, when a voltage of 2.3~V was applied to the heater. To measure the switching time between these two wavelengths, the switching is induced by applying a voltage step from 0~V to 2.3~V to the heater. To observe the spectral response of the laser, the output is normally incident a reflection grating (600~lines/mm). A photodiode is placed in one of the two transverse positions, 2~m from the grating, corresponding to diffraction of light from either $\lambda$=1552.2~nm or $\lambda$=1548.2~nm. Fig.~\ref{FigTuningSpeed} shows the photodiode signal recorded as a function of time. The traces show a step-like shape, from which the switching time can be retrieved. The time interval in which the signal changes from 10\% to 90\%, from 1552.2~nm to 1548.2~nm and vice versa can be seen to be 200~$\mu$s~$\pm$40~$\mu$s.

To summarize, our measurements show that, compared to the free-running laser diode, the external cavity improves the SMSR by 30~dB (from 20~dB to 50~dB). The laser's tuning range is measured to be 44~nm and the switching time between two wavelengths is 200~$\mu$s. The laser bandwidth is as narrow as 25~kHz.

\section{Conclusion}
\label{Conclusion}
In this letter we have investigated the spectral properties of a waveguide based external cavity semiconductor laser (WECSL), where the external cavity is a tunable micro-ring resonator mirror, fabricated in box-shaped Si$_3$N$_4$/SiO$_2$ waveguide technology. The SMSR of the WECSL is measured to be 50~dB. Our laser is tunable over the full telecommunications C-band and switches between two preset wavelengths in 200~$\mu$s. Most importantly, our WECSL offers an extremely small spectral bandwidth of 25~kHz. The fundamental (Schawlow-Townes) limit, reduced by the MRR-based enhancement of the cavity photon lifetime and reduced by the spectral narrowing of the external feedback, is in the order of only 5~Hz. This indicates that, via technical improvements it should be possible to reduce the experimental bandwidth of the laser by several orders of magnitude. The narrow bandwidth and great potential for significant further narrowing, fast wavelength switching and wide wavelength tunability make this WECSL an ideal candidate for advanced applications such as in broadband squint-free K$_u$-band phased array antenna systems.


\end{document}